\documentclass[12pt,preprint]{aastex}

\def\cs{{c_{\rm s}}}
\newcommand{\dfrac}[2]{\frac{\displaystyle{#1}}{\displaystyle{#2}}}
\newcommand{\bld}[1]{\mbox{\boldmath$#1$}}

\slugcomment{Accepted to ApJ}

\shorttitle{Dust Evolution in a Protoplanetary Disk}
\shortauthors{Nomura \& Nakagawa}

\begin{document}


\title{Dust Size Growth and Settling in a Protoplanetary Disk}


\author{Hideko Nomura and Yoshitsugu Nakagawa}
\affil{Department of Earth and Planetary Sciences, Kobe University,
Kobe 657-8501, Japan}
\email{hnomura@kobe-u.ac.jp, yoshi@kobe-u.ac.jp}




\begin{abstract}
We have studied dust evolution in a quiescent or turbulent
 protoplanetary disk by numerically solving coagulation equation for
 settling dust particles, using the minimum mass solar nebular model.
 As a result, if we assume an ideally quiescent disk, the dust
 particles  settle toward the disk midplane to form a gravitationally
 unstable layer within $2\times 10^3$--$4\times 10^4$yr at 1--30 AU,
 which is in good agreement with an analytic calculation by Nakagawa,
 Sekiya, \& Hayashi (1986) 
 although they did not take into account the particle size distribution
 explicitly. In an opposite extreme case of a globally
 turbulent disk, on the other hand, the dust particles
 fluctuate owing to turbulent motion of the gas
 and most particles become large enough to move
 inward very rapidly within 70--$3\times 10^4$yr at 1--30 AU,
 depending on the strength of turbulence. Our result
 suggests that global turbulent motion should cease for the
 planetesimal formation in protoplanetary disks. 
\end{abstract}





\section{Introduction}

It is believed that particle settling and growth are important processes
leading to the planet formation in protoplanetary disks. 
Observationally some evidences of dust size growth have been proposed based
on dust continuum emission from the protoplanetary disks, such as smaller
power-law indices of spectral energy distributions (SEDs) in sub-mm to
cm wavelength bands, and fainter trapezium feature of 10$\micron$
silicate emission, compared with those of the interstellar dust grains
(e.g., Beckwith \& Sargent 1991; Miyake \& Nakagawa 1993;
Beckwith et al. 2000; Kitamura et al. 2002; van Boekel et al. 2003;
Przygodda et al. 2003; Wilner et al. 2005).
Meanwhile, theoretically the initial stage of dust evolution
when micron-sized interstellar dust particles grow into centimeter-sized
particles with settling toward the disk midplane have been studied
analytically and numerically; for example,
Weidenschilling (1980) and Nakagawa et al. (1981, 1986)'s works in a
quiescent disk, and Cuzzi et al. (1993) and Weidenschilling (1997,
2004)'s works in local turbulence induced by the shear between the dust
layer and the gas near the disk midplane
(see also Weidenschilling \& Cuzzi 1993 and references therein).

In addition to the shear-induced turbulence, it is thought that there
exists global turbulent motion in the protoplanetary disks, which is
caused by thermal convective and/or magneto-rotational instabilities
(e.g., Lin \& Papaloizou 1980; Balbus \& Hawley 1991). The turbulent
motion of the gas is known to affect the dust evolution processes;
turbulence induced relative motion increases the mutual collision rate,
that is, the growth rate of the dust particles (e.g., 
V\"{o}lk et al. 1980), turbulent mixing motion lets the particles move
diffusively (e.g., Cuzzi et al. 1993), and turbulent eddies trap the
dust particles (e.g., Klahr \& Henning 1997; Cuzzi et al. 2001;
Johansen et al. 2004).
On the other hand, the disk instabilities, namely, the existence of
turbulent regions depend on the spatial and size distributions of the
dust particles (e.g., Mizuno et al. 1988; Sano et al. 2000; Nomura 2004).
Therefore, self-consistent treatment of the evolution of the dust
particles and the
turbulent regions (the disk instabilities) is needed in order to
understand the very beginning of planet formation process in
protoplanetary disks before the dust particles settle toward the disk
midplane and form a dusty layer, which could lead to the planetesimal
formation. 

Moreover, recent observations have been providing a huge amount of data
of spectra and SEDs of dust continuum emission as well as molecular line
emissions from protoplanetary disks. Theoretically reproducing these
observational data have been also developed using detailed disk models
(e.g., Kenyon \& Hartmann 1987; Miyake \& Nakagawa 1995; Chiang 
\& Goldreich 1997; D'Alessio et al. 1998; Dullemond et al. 2001;
van Zadelhoff et al. 2001; Aikawa et al. 2002; Nomura \& Millar 2005).
Although rather simple dust models have been used in those previous models,
the dust properties affect the physical and chemical structure of the
disks, and then the observable properties very much (e.g., D'Alessio et
al. 2001; Aikawa \& Nomura 2005). Thus, we should carefully model the
dust evolution in the disks and compare the model predictions with the
observational data in order to interpret the observations and understand 
what is actually going on in protoplanetary disks.
The SEDs of young stellar objects have been tried to be modeled
by numerically simulating the dust size growth and settling
processes in the disks (Suttner \& Yorke 2001; Tanaka et al. 2005;
Dullemond \& Dominik 2005).

In this paper, we have studied basic behavior of dust evolution in a
quiescent or globally turbulent protoplanetary disk, especially focusing
on time scales and growing dust size. It is done by numerically solving
the coagulation equation, using a simple disk model, as a first step 
of understanding the initial stage of planet formation and modeling
observational properties of protoplanetary disks. In the following
section, we present the disk model as well as the basic equations for
the vertical and radial motion and the coagulation of dust particles. In
\S 3, we numerically calculate the dust size growth and settling toward
the midplane in a quiescent disk, and compare the result with that
obtained analytically by Nakagawa  
et al. (1986; hereafter NSH86). We also discuss the dust evolution in a
globally turbulent disk in \S 4. Finally, the results are summarized in
\S 5.

\section{Basic Equations and Models}

\subsection{Disk Model}

As a disk model, we adopt the minimum mass solar nebular model (e.g.,
Safronov 1969; Hayashi 1981; Hayashi et al. 1985) in order to examine
basic behavior of
the dust evolution under a simple physical condition of the gas and
compare the results with analytic calculation (see \S \ref{Sect3}). In
this model the gas surface density profile is given by
\begin{equation}
\Sigma_{\rm gas}=1.7\times 10^3(R/{\rm AU})^{-3/2} {\rm g\ cm}^{-2},
\end{equation}
where $R$ is the radial distance from the central star.
The surface density of dust particles is 
\begin{equation}
\Sigma_{\rm dust}=\left\{\begin{array}{l}
7.1 \\
30
\end{array}\right\}
(R/{\rm AU})^{-3/2} {\rm g\ cm}^{-2}\ \ \ 
{\rm for}\left\{\begin{array}{l}
R<2.7{\rm AU} \\
R>2.7{\rm AU}
\end{array}\right., \label{eq2}
\end{equation}
where the solid density of a dust particle, $\rho_s$, is set to be
$\rho_s=2$ and 1 g cm$^{-3}$ for $R<2.7$ {\rm AU} and $R>2.7$ {\rm AU},
respectively, taking into account the effect of water ice sublimation.
We note that each of the mass and the surface density distribution
is one of the most unknown factors in modeling protoplanetary
disks. Numerical calculation of dust evolution using some different
parameters for the disk mass and the power-low index of the surface
density distribution as a function of the radial distance is performed
by Tanaka et al. (2005).
The dust and gas temperature is given by
\begin{equation}
T=280(R/{\rm AU})^{-1/2} {\rm K}. \label{eq3}
\end{equation}
In this paper we treat the region where the vertical distance from the
disk midplane, $Z$, is smaller than the disk scale height,
\begin{equation}
H=(\sqrt{\pi}/2)(\cs/\Omega_{\rm K})=0.0472(R/{\rm AU})^{5/4}{\rm AU},
\end{equation}
where $\cs=(8kT/\pi m_{\mu})^{1/2}$ ($m_{\mu}=2.34m_{\rm H}$ is the mean
molecular mass and $m_{\rm H}$ the mass of an atomic hydrogen) is the
mean thermal velocity and $\Omega_{\rm K}=(GM_*/R^3)^{1/2}$ ($G$ is the
gravitational constant and $M_*$ the mass of the central star) is the
Keplerian frequency. And we simply assume that the profiles of
the temperature (eq. [\ref{eq3}]) and the gas density, 
\begin{equation}
\rho_{\rm gas}=\Sigma_{\rm gas}/\sqrt{\pi} H=1.36\times 10^{-9}(R/{\rm AU})^{-11/4} {\rm g\ cm}^{-3},
\end{equation}
are uniform in the vertical direction. The mass of the central star is
assumed to be $M_*=1M_{\odot}$.

\subsection{Equations of Motion of Dust Particles and Gas}

The equations of motion of dust particles and gas in the disk are given by
\begin{equation}
\dfrac{d\bld{U}}{dt}=-A\rho_{\rm gas}(\bld{U}-\bld{u})-\dfrac{GM_*}{R^3}\bld{R}, \label{eq6}
\end{equation}
and
\begin{equation}
\dfrac{d\bld{u}}{dt}=-A\rho_{\rm dust}(\bld{u}-\bld{U})-\dfrac{GM_*}{R^3}\bld{R}-\dfrac{\nabla p_{\rm gas}}{\rho_{\rm gas}},
\end{equation}
where $\bld{U}$ and $\bld{u}$ are the velocities of dust and gas
particles in the inertial frame of reference, and $\rho_{\rm dust}$ the
spatial mass density of dust particles, 
$p_{\rm gas}$ the gas pressure. The drag coefficient, $A$, is given by
\begin{equation}
A=\left\{\begin{array}{ll}
\cs/\rho_sa & {\rm for}\ \ a\la l_g, \\
3\cs l_g/2\rho_sa^2 & {\rm for}\ \ a\ga l_g,
\end{array}\right.  \label{eq8}
\end{equation}
following Epstein's and Stokes' law (Epstein 1924; Stokes 1851),
respectively. The symbol $a$ is the radius of the dust particles
and $l_g$ is the mean free path of the gas
particles, given by $l_g=m_{\mu}/(\sigma_{\rm mol}\rho_{\rm
gas})=1.44(R/{\rm AU})^{11/4}$ cm, where $\sigma_{\rm mol}=2\times
10^{-15}$ cm$^2$ is the molecular cross section. 
The shape of the dust particles is simply assumed to be a compact sphere
in this paper. We note that the dust shape (mass/area ratio)
affects the dust 
evolution through the drag coefficient and the sticking rate (see next
subsection), and many numerical studies have dealt with the effects of
the dust shape, taking into account fractal structure of dust aggregates
(e.g., Weidenschilling \& Cuzzi 1993; Ossenkopf 1993; Weidenschilling
1997; Suttner \& Yorke 2001; Dullemond \& Dominik 2005).

Now, we assume that the disk is axisymmetric and rotates around the
central star at nearly Keplerian velocity, and
set the dust and gas velocities 
relative to the Keplerian velocity, $\bld{v}_{\rm K}(=R\Omega_{\rm
K}\bld{\hat{\phi}})$, as
$\bld{V}=\bld{U}-\bld{v}_{\rm K}$ and $\bld{v}=\bld{u}-\bld{v}_{\rm K}$.
As far
as the dust particles are small enough, the timescale for initial
velocity of dust particles decaying due to the gas drag force is 
much shorter than the Keplerian time and the timescale of collision
between dust particles (e.g., Nakagawa et al. 1981). Thus, the mean
motion of dust particles becomes steady soon in a quiescent disk (see
also \S \ref{S4.1}). 
By setting $\partial /\partial t=0$ in the equations of motion, we can
derive the terminal velocities of the vertical and radial motion of dust
particles as
\begin{equation}
V_Z-v_Z=-(\Omega_{\rm K}^2/D_g)Z, \label{eq9}
\end{equation}
where $D_g=A\rho_{\rm gas}$, and
\begin{equation}
V_R-v_R=-\dfrac{2D\Omega_{\rm K}}{D^2+\Omega_{\rm K}^2}\eta v_{\rm K} \ \ \ {\rm and}\ \ \ V_R=-\dfrac{\rho_{\rm gas}}{\rho_{\rm gas}+\rho_{\rm dust}}\dfrac{2D\Omega_{\rm K}}{D^2+\Omega_{\rm K}^2}\eta v_{\rm K}, \label{eq10}
\end{equation}
where $D=A(\rho_{\rm gas}+\rho_{\rm dust})$ and
\begin{equation}
\eta=-\dfrac{1}{2R\Omega_{\rm K}^2}\dfrac{1}{\rho_{\rm gas}}\dfrac{\partial p_{\rm gas}}{\partial R}=1.81\times 10^{-3}(R/{\rm AU})^{1/2}
\end{equation}
(see NSH86 for details).

\subsection{Coagulation Equation for Settling Particles} \label{Sect2.3}

We solve the following dispersed coagulation equation numerically for
simulating the size growth of settling dust particles in the disk,
according to Nakagawa et al. (1981) and Nakagawa \& Kohno (1999);
\begin{equation}
\dfrac{\partial\varphi(i)}{\partial t}+\dfrac{\partial}{\partial Z}[V_Z(i)\varphi(i)]=-m_i\varphi(i)\sum_{j=1}^n\beta(i,j)\varphi(j)+\dfrac{1}{2}m_i\sum_{j=1}^{i-1}\beta(i-j,j)\varphi(i-j)\varphi(j),  \label{eq12}
\end{equation}
where
\begin{equation}
\varphi(i)=\int_{m_{i-1/2}}^{m_{i+1/2}}\rho(m)dm
\end{equation}
is the mass density of the dust particles whose mass ranges from
$m_{i-1/2}$ to $m_{i+1/2}$, and $m_i=(m_{i-1/2}+m_{i+1/2})/2$ for
$i=1,\cdots,n$. The dust mass is binned into $n$ intervals
logarithmically as $m_{i+1/2}=\varepsilon m_{i-1/2}$, where 
$m_i=(4\pi/3)\rho_sa_i^3$ (the dust particles are assumed to have a
shape of compact sphere), $a_1=1\micron$, $n=320$, and
$\varepsilon=\sqrt[4]{2}$
are adopted in this paper (see Appendix). Here, the total mass density
of dust particles at a given position and time is given by
\begin{equation}
\rho_{\rm dust}=\int_{m_{1/2}}^{m_{n+1/2}}\rho(m)dm=\sum_{i=1}^n\varphi(i).
\label{eq14}
\end{equation}

The second term of the left hand side of equation (\ref{eq12}) shows the
mass transport of dust particles in the vertical direction. 
Now, the mean vertical velocity of the gas is negligible
($v_z\approx 0$), so the dust particles settle toward the disk midplane.
The mean settling velocity of dust particles with mass $m_i$ is derived
as $V_Z(i)=-(\rho_{\rm gas}\cs/\rho_sa_i)\Omega_{\rm K}^2Z$ (if $a_i\la
l_g$) or $-(3\rho_{\rm gas}\cs l_g/2\rho_sa_i^2)\Omega_{\rm K}^2Z$
(if $a_i\ga l_g$) from equations (\ref{eq8}) and (\ref{eq9}). In a
turbulent disk
the dust particles are transported by turbulent mixing in addition (see
\S \ref{Sect4.3}). The mass transport of dust particles in the radial
direction is not solved in this paper for simplicity.

The symbol $\beta(i,j)$ is related to the sticking rate of two colliding
dust particles, and given by
\begin{equation}
\beta(i,j)=\pi(a_i+a_j)^2\delta Vp_{\rm s}/m_im_j, \label{eq15}
\end{equation}
where we simply assume the sticking probability of $p_{\rm s}=1$ in this
paper. We note that the sticking probability will depend on
size, relative velocities, chemical composition and/or shape of
dust grains (e.g., Weidenschilling \& Cuzzi 1993; Weidenschilling 2004;
see also references therein). Lower probability will make the timescale of
the dust evolution longer
(e.g., Tanaka et al. 2005). Here, we neglect fragmentation of
dust particles, which could occur if the particles become large and
their relative velocities with small particles become high enough (e.g.,
Dullemond \& Dominik 2005).
As the relative velocity between the dust particles, $\delta V$,
we take into account the thermal Brownian motion,
\begin{equation}
\delta V_{\rm B}=\biggl(\dfrac{8kT}{\pi}\biggr)^{1/2}\biggl(\dfrac{1}{m_i}+\dfrac{1}{m_j}\biggr)^{1/2},
\end{equation}
where $k$ is Boltzmann's constant, and the velocity differences in the
vertical and radial directions, $\delta V_Z=V_Z(i)-V_Z(j)$ and $\delta
V_R=V_R(i)-V_R(j)$, which are derived from equations (\ref{eq9}) and
(\ref{eq10}), respectively. The azimuthal velocity difference, $\delta
V_{\phi}$, has very weak size dependence as far as the dust particles
are small (e.g., NSH86); hence we neglect $\delta V_{\phi}$.
We adopt the relative velocity of $\delta
V=(\delta V_{\rm B}^2+\delta V_Z^2+\delta V_R^2)^{1/2}$ in a quiescent
disk. In a turbulent disk we take into account the turbulence induced
relative velocity, $\delta V_t$, in addition as $\delta
V=(\delta V_{\rm B}^2+\delta V_Z^2+\delta V_R^2+\delta V_t^2)^{1/2}$
(see \S \ref{Sect4.3}).


In our numerical calculation the spatial grid in the vertical direction
is taken equally spaced into 20 intervals within $0<Z<H$. In addition,
within the lowest interval of $0<Z<H/20$, we take logarithmically spaced
17 sub-intervals as $Z_{l+1}=2Z_l$ in order to resolve the region near
the disk midplane in a quiescent disk. In a turbulent disk only equally
spaced 20 intervals within $0<Z<H$ are used without sub-intervals
because the dust size distributions 
are almost identical between the lowest vertical disk layers and the
calculation of the diffusive mass transport due to the turbulent mixing
is very time consuming if we use such small spatial sub-grids
(see \S\ref{Sect4.3}).

\section{Dust Evolution in a Quiescent Disk} \label{Sect3}

By numerically solving the coagulation equation for settling dust
particles (eq. [\ref{eq12}]), we obtain the dust size distributions at
a given time and disk height at the Earth's (1AU),
Jupiter's (5.2AU), or Neptune's (30AU) orbits. As an
initial condition, we assume that the dust particles are well-mixed with
the gas and have a radius of a certain value, $a_{\rm init}$. We adopt
$a_{\rm init}=10$, 20, and 60 $\micron$ for $R=1$, 5.2, and 30 AU,
respectively, in order to compare the numerical results with 
the analytic calculation by NSH86. These values correspond to the
wavelength of the peak emission at the local temperature and do not have
particular physical meaning. The initial condition of the dust
size distribution, however, does not affect the results very much. 

\begin{figure}
\begin{center}
\includegraphics[scale=1.0]{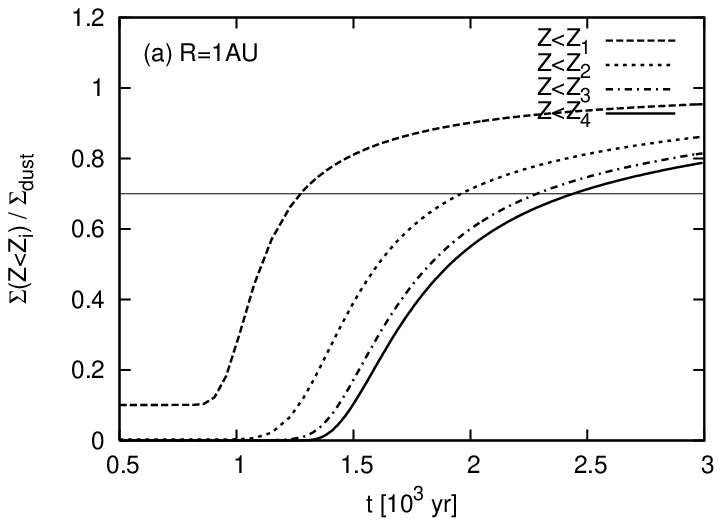}

\includegraphics[scale=1.0]{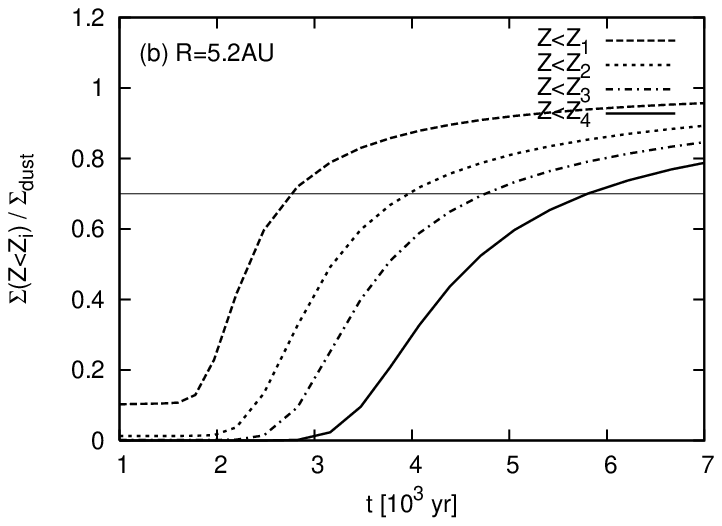}

\includegraphics[scale=1.0]{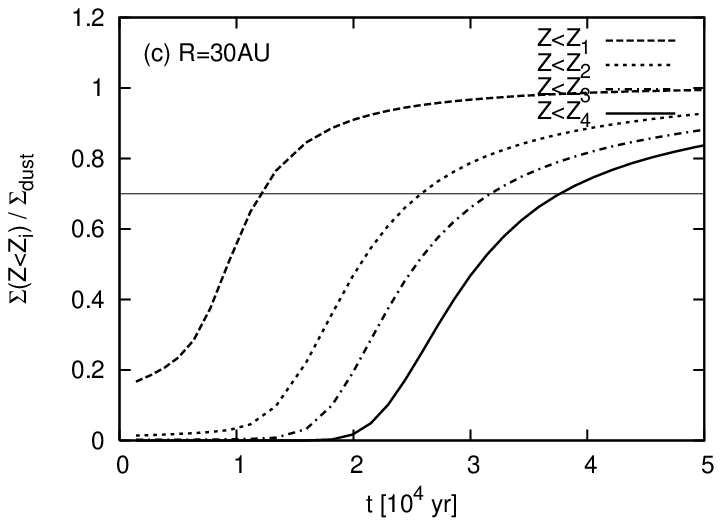}
\caption{The time evolution of spatial dust mass distribution obtained
 by our numerical calculation at the orbits of $R=$ (a) 1AU, (b)
 5.2AU, and (c) 30AU. The dashed, dotted, dot-dashed, and solid lines
 represent the dust surface density below the characteristic heights,
 $Z=Z_1, Z_2, Z_3$, and $Z_4$, 
 respectively, divided by the total dust surface density, $\Sigma_{\rm dust}$.
 The thin solid lines at $\Sigma(Z<Z_i)/\Sigma_{\rm dust}=0.7$ are used
 in order to estimate the dust settling time in Figure~\ref{f3}. 
 \label{f2}} 
\end{center}
\end{figure}

\begin{figure}
\plotone{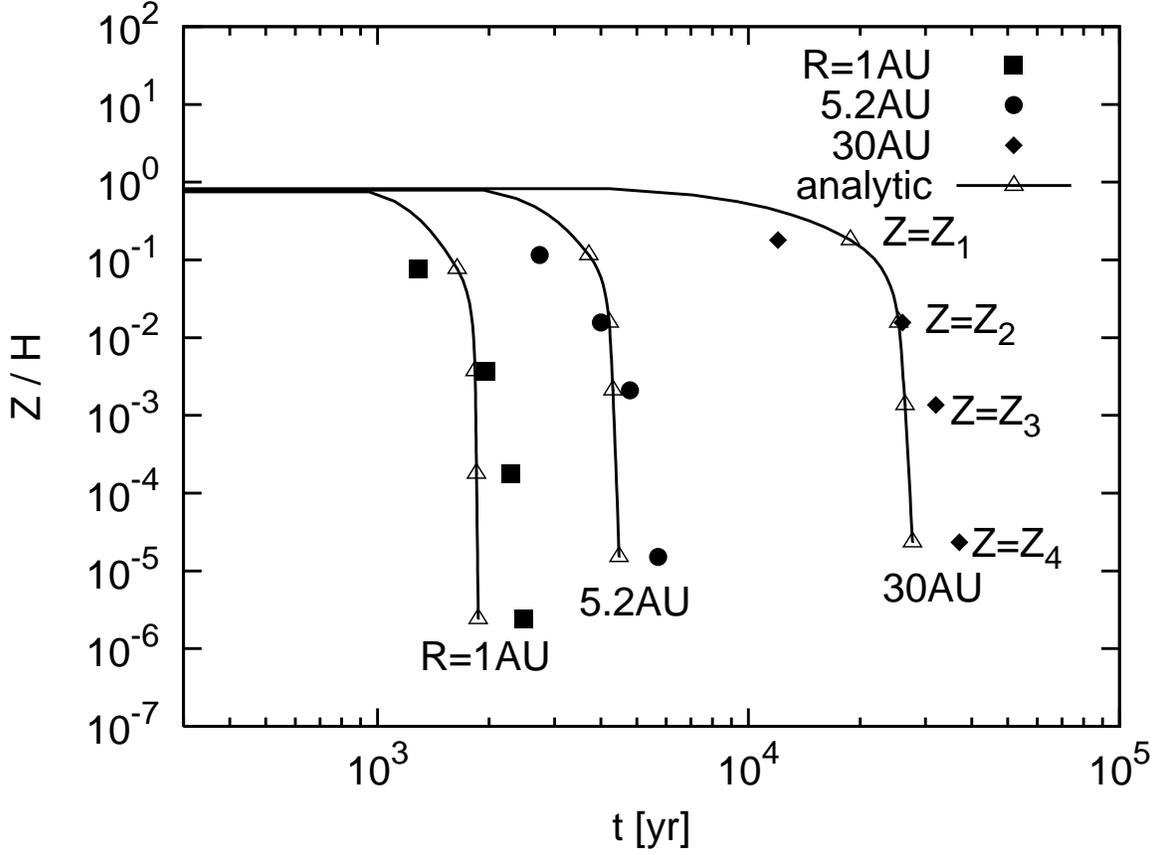}
\caption{The dust settling time at the characteristic height $Z=Z_i$
 ($i=1,\cdots,4$) and $R=1$AU 
 (squares), 5.2AU (circles), and 30AU (diamonds), obtained by our
 numerical calculation. The dust settling time by NSH86 is also plotted
 (triangles with solid lines). The numerical results are in good
 agreement with the analytic results, and most dust particles settle
 below $Z=Z_4$, within about $2\times 10^3$, $5\times 10^3$, and
 $3\times 10^4$ yrs at the orbits of $R=$1, 5.2, and 30 AU, respectively,
 in a quiescent disk. \label{f3}} 
\end{figure}

First, we compare our numerical result of the dust settling time with
that by NSH86. Figure \ref{f2} shows the time evolution of spatial dust
mass distribution obtained by our numerical calculation at the orbits
of $R=$ (a) 1AU, (b) 5.2AU, and (c) 30AU. The vertical
axis represents the dust surface density from $Z=0$ to a characteristic
height $Z=Z_k$ ($k=1,\cdots,4$) at $R$,
$\Sigma(Z<Z_k)=\int_{-Z_k}^{Z_k}\rho_{\rm dust}(R,Z)dZ$, divided by the
total dust surface density there, $\Sigma_{\rm dust}$ (eq.[\ref{eq2}]). 
The values, $Z_1, Z_2, Z_3$, and $Z_4$, at $R=$1, 5.2, 
and 30 AU are listed in Table \ref{T1}. These characteristic
heights are defined in NSH86 (see Fig.1 of NSH86); to put it briefly,
the vertical velocity, $V_Z$ (eq. [\ref{eq9}]), of dust particles
dominates the radial velocity, $V_R$ (eq. [\ref{eq10}]), above $Z=Z_1$,
and the gas density, $\rho_{\rm gas}$, is larger than the dust density,
$\rho_{\rm dust}$, above $Z=Z_2$. The vertical velocity, $V_Z$,
dominates the radial velocity, $V_R$, again 
below $Z=Z_3$ where the dust density is high enough that the gas drag
force hardly affects the radial motion of the dust particles. If most
dust particles settle below $Z=Z_4$, the dust layer becomes
gravitationally unstable and could fragment into planetesimals. 
The dashed, dotted, dot-dashed, and solid lines in Figure \ref{f2}
represent the dust surface density below 
$Z=Z_1, Z_2, Z_3$, and $Z_4$, respectively. As time goes on, the dust
particles settle toward the disk midplane and more mass is included in
the lower layer of the disk. In Figure \ref{f3} we plot the dust
settling time at which 70\% of the total dust mass settles below $Z_k$
($k=1,\cdots,4$) at the orbits of $R=1$AU (squares), 5.2AU (circles), and
30AU (diamonds). Together with them, the dust settling time obtained by
NSH86 is also plotted (triangles with solid lines). The 
figure shows that the numerical results are in good agreement with the
analytic results, and most dust particles settle below $Z=Z_4$,
which leads to the formation of a gravitationally unstable dust layer,
within about $2\times 10^3$, $6\times 10^3$, and $4\times 10^4$ yrs at
the orbits of $R=$1, 5.2, and 30 AU, respectively, in a quiescent disk.

\begin{figure} 
\includegraphics[scale=1.0]{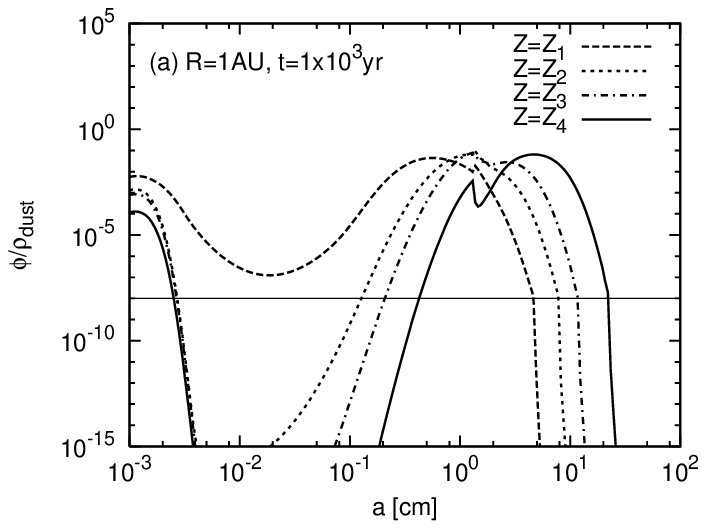}
\includegraphics[scale=1.0]{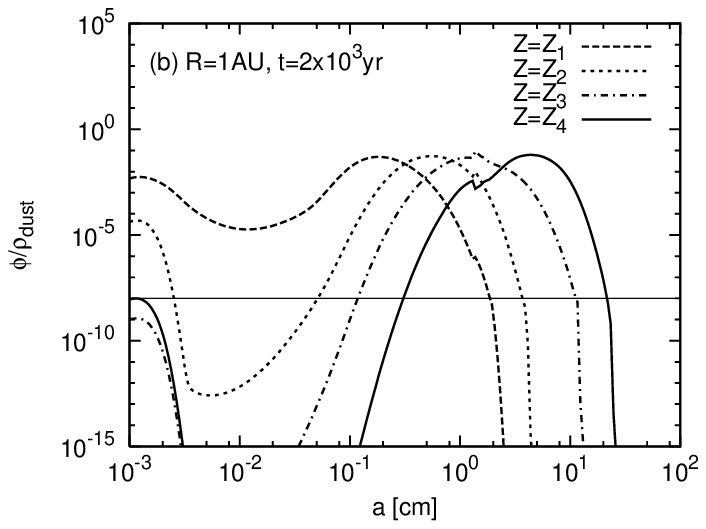}

\includegraphics[scale=1.0]{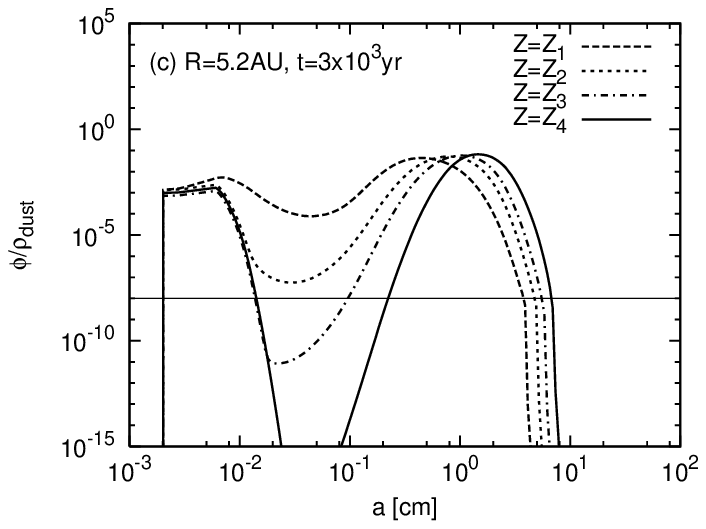}
\includegraphics[scale=1.0]{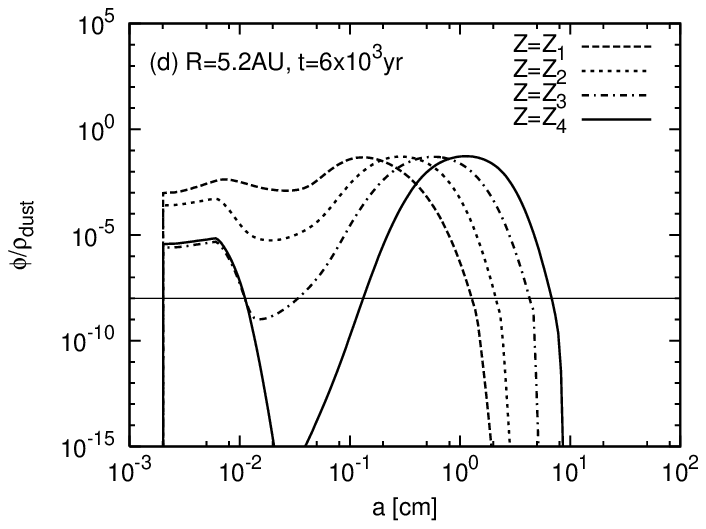}

\includegraphics[scale=1.0]{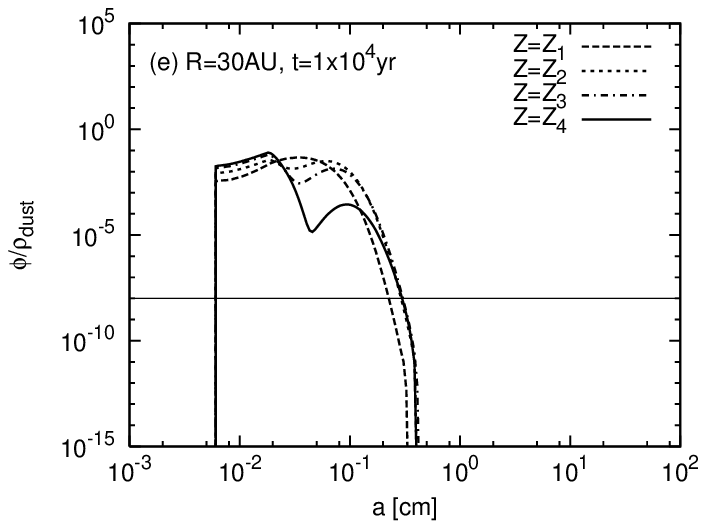}
\includegraphics[scale=1.0]{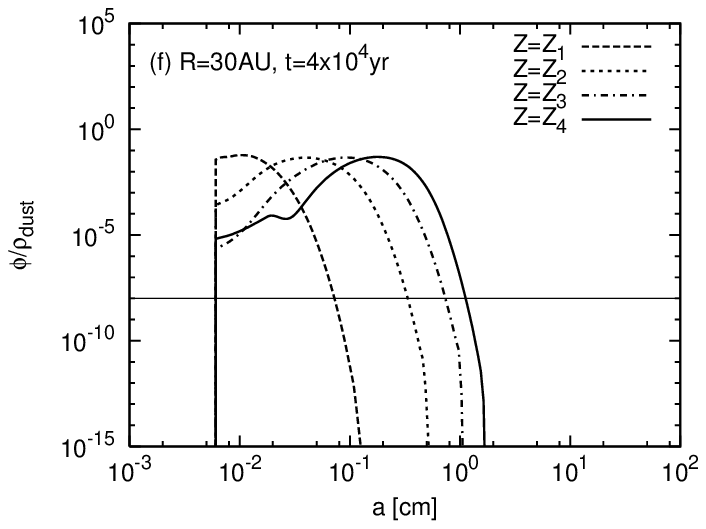}
\caption{The size distributions of mass density of dust particles,
 $\varphi(i)$, normalized by $\rho_{\rm dust}$ at each orbit, $R$,
 and time, $t$, in a quiescent disk. The dashed, dotted, dot-dashed, and
 solid lines represent the size distributions at the characteristic
 heights $Z=Z_1, Z_2, Z_3$, and $Z_4$, 
 respectively. The thin solid lines at $\varphi/\rho_{\rm dust}=10^{-8}$
 are used as a criterion to estimate the largest dust radii in Figure
 \ref{f4}. \label{f1}} 
\end{figure}

Next, Figure \ref{f1} shows the resulting size distributions of mass
density of dust particles, $\varphi(i)$, normalized by $\rho_{\rm dust}$
at (a) $R=1$AU, $t=1\times 10^3$ yr, (b) $R=1$AU, $t=2\times 10^3$ yr;
(c) $R=5.2$AU, $t=3\times 10^3$ yr, (d) $R=5.2$AU, $t=6\times 10^3$ yr;
(e) $R=30$AU, $t=1\times 10^4$ yr, and (f) $R=30$AU, $t=4\times 10^4$ yr.
The dashed, dotted, dot-dashed, and solid lines in each figure represent
the size distributions at the characteristic heights, $Z=Z_1, Z_2, Z_3$,
and $Z_4$, respectively. 
The time used in Figure \ref{f1}{\it a, c}, and {\it e} corresponds to
$t=t_1$ and that in Figure \ref{f1}{\it b, d}, and {\it f} corresponds
to $t=t_4$, where $t_k$ denotes the time at which 70\% of the dust
particles settle below the characteristic height, $Z=Z_k$, at each
orbit, $R$ (see Fig.~\ref{f3}). 
Now, we can
see from Figure \ref{f1} that the larger dust particles which have grown
at the disk surface layer settle more rapidly toward the midplane with
growing larger and larger. These processes lead to a bimodal size
distribution 
near the midplane at the inner disk (e.g., Weidenschilling 1997). The
gaps appear in Figure \ref{f1}{\it a} and {\it b} around $a=l_g=1$ cm
where the drag coefficient, $A$ (eq.[\ref{eq8}]), begins to follow
Stokes' law, rather 
than Epstein's law. The drag coefficient always follows Epstein's law
at $R=5.2$ and 30AU in this model, where $l_g$ is much larger
($l_g=1\times 10^2$ and $2\times 10^4$ cm, respectively).
The timescale of the dust size growth and settling is shorter at
the inner disk where the effect of gravitational force of the
central star is stronger.

\begin{figure}
\plotone{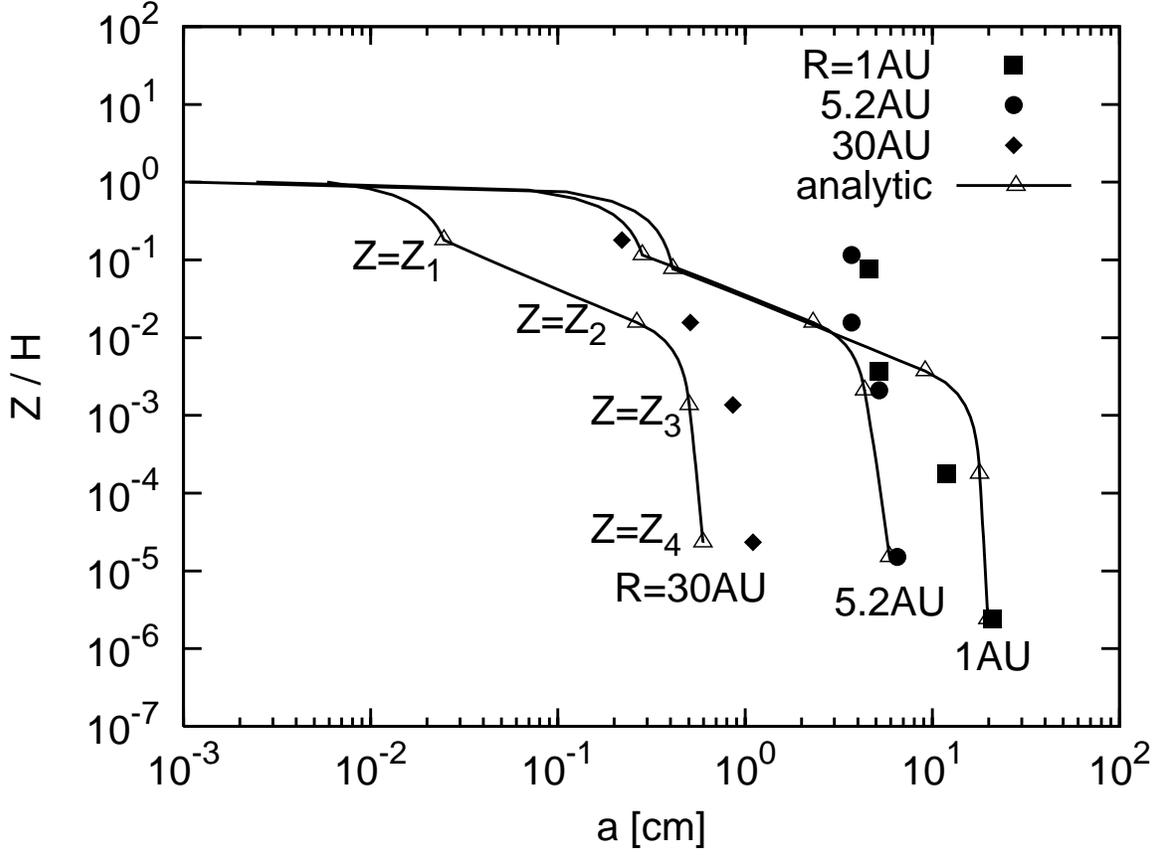}
\caption{The largest dust radii at the characteristic height $Z=Z_i$
 ($i=1,\cdots,4$) and $R=1$AU 
 (squares), 5.2AU (circles), and 30AU (diamonds), obtained by our
 numerical calculation. The evolution of dust radii by NSH86 is also
 plotted (triangles with solid lines). The numerical results are in
 good agreement with the analytic results within a factor of two,
 except at $Z=Z_1$, and the dust radius finally reaches
 about 20, 7, and 1 cm at the orbits of $R=$1, 5.2, and 30 AU, 
 respectively, just before most dust particles settle below $Z=Z_4$.
 \label{f4}} 
\end{figure}

Finally, we compare our numerical result of the evolution of dust particle
radius with that by NSH86. In Figure \ref{f4} we plot the resulting
largest dust radii at $Z=Z_k$ ($k=1,\cdots,4$) and $R=1$AU (squares),
5.2AU (circles), and 30AU (diamonds). The largest dust radii, $a_{\rm
max}$, at $Z=Z_k$ are obtained applying a criterion, $i_{\rm
max}=\max\{\ i\ |\ \varphi(i)/\rho_{\rm dust}>10^{-8}\ \}$, to the dust
size distribution at $t=t_k$ (cf. thin solid lines in Fig.~\ref{f1}).
The evolution of dust radii obtained by NSH86 is also plotted (triangles
with solid lines). The figure shows that the numerical results are in
good agreement with the analytic results within a factor of two,
except at $Z=Z_1$. The dust radii at $Z=Z_1$ are larger in the numerical
calculation because the relative velocity between dust particles due to
the thermal Brownian motion, which works efficiently for small dust
particles, is taken into account in the
numerical calculation, but not in the analytic calculation in NSH86.
Both numerical and analytic
calculations show that the dust particles grow and their radii finally
reach about 20, 7, and 1 cm at the orbits of $R=$1, 5.2, and 30 AU,
respectively, just before they settle below $Z=Z_4$.

We note that the orbital decay is very little in a quiescent disk;
during settling from $Z=H$ to $Z=Z_4$, the dust particles move radially 
by $\Delta R=2.2\times 10^{-3}$, 0.20, and 2.8AU at the orbits
of $R=1$, 5.2, and 30AU, respectively, according to NSH86. 

Here it should be commented that although we simply assume a totally
quiescent disk in this section, in reality it is expected that the shear
between the dust layer and the gas induces turbulence locally near the
midplane as the dust particles settle (e.g., Cuzzi et al. 1993;
Weidenschilling \& Cuzzi 1993; Weidenschilling 1997; Cuzzi \&
Weidenschilling 2005). Actually if we compute the 
Richardson number, $J=-(\partial \rho_{\rm dust}/\partial Z)(\rho_{\rm
gas}+\rho_{\rm dust})^{-1}\Omega_{\rm K}^2Z(\partial V_{\phi}/\partial
Z)^{-2}=Z(\eta R\rho_{\rm gas})^{-2}(\rho_{\rm gas}+\rho_{\rm
dust})^3(\partial \rho_{\rm dust}/\partial Z)^{-1}$ (e.g., Sekiya 1998;
Chandrasekhar 1961), we can find that 
$J<0.25$ and the shear instability will occur below $Z=Z_2$ at each 
orbit. In such a turbulent layer the dust particles do not concentrate
in the midplane and migrate inward very rapidly once they grow large
enough as we will see in next section (although it is different in point
that this is local turbulence). 

In conclusion, our numerical calculation have confirmed that NSH86's
approximate treatment of the dust size growth and settling processes are
appropriate for describing the dust settling time and the evolution of
the largest dust size in an ideally 
quiescent disk, although they did not take
into account the dust size distribution explicitly. This will be because
most dust mass is included in the dust particles with the 
largest sizes, and the dust settling time is controlled by the largest
dust particles which have the highest settling velocity.


\section{Dust Evolution in a Turbulent Disk} \label{S4}

In this section we discuss the dust evolution in the disk in which
global turbulent motion exists induced by, for example, thermal
convective and/or magneto-rotational instabilities.

\subsection{Vertical Motion}\label{S4.1}

First, we examine the vertical motion of one dust particle, not the
mean motion which we have treated in the previous sections. The
equation of motion of a dust particle (eq.[\ref{eq6}]) in the vertical
direction is written as 
\begin{equation}
\dfrac{d^2Z}{dt^2}=-D_g\biggl(\dfrac{dZ}{dt}-u_z\biggr)-\Omega_{\rm K}^2Z. \label{eq17}
\end{equation}
The vertical velocity of the gas in a turbulent medium is generally
given by $u_z=\overline{u_z}+u_z'$, where the overline means a time
average and 
the prime is a fluctuation due to the turbulent motion. Now we put the
mean velocity to be $\overline{u_z}=0$ since we assume the hydrostatic
equilibrium. As the component of turbulent fluctuation,
we simply adopt an oscillating motion of $u_z'=v_t\exp(i\omega_tt)$,
which models the motion of the largest turbulent eddy with a velocity
$v_t$ and a frequency $\omega_t$ ($\omega_t\sim v_t/l_t$ where $l_t$ is
the eddy size). In this case, equation (\ref{eq17}) has 
a general solution that consists of a mean motion part,
$\overline{Z}(t)$, and a fluctuation part, $Z'(t)$, caused by the
turbulent motion of the gas,
\begin{equation}
Z(t)=\overline{Z}(t)+Z'(t),
\label{eq19.0}
\end{equation}
where
\begin{equation}
\overline{Z}(t)=C_1\exp(\lambda_1t)+C_2\exp(\lambda_2t),
\label{eq19.1}
\end{equation}
and
\begin{equation}
Z'(t)=\dfrac{D_gv_t\exp[i(\omega_{\rm t}t-\delta)]}{[(\Omega_{\rm K}^2-\omega_{\rm t}^2)^2+D_g^2\omega_{\rm t}^2]^{1/2}},\ \ \ \delta=\tan^{-1}\dfrac{D_g\omega_{\rm t}}{\Omega_{\rm K}^2-\omega_{\rm t}^2}.
\label{eq19.2}
\end{equation}
In equation (\ref{eq19.1})
$C_1$ and $C_2$ are integral constants, and
\begin{eqnarray}
\lambda_{1,2} & = & -\dfrac{1}{2}(D_g\pm\sqrt{D_g^2-4\Omega_{\rm K}^2}) \\
 & \simeq & \left\{\begin{array}{ll}
     -D_g,\ -\Omega_K^2/D_g & {\rm for}\ \ D_g\gg 2\Omega_{\rm K}, \\
     -(D_g/2)\mp i\Omega_{\rm K} & {\rm for}\ \ D_g\ll 2\Omega_{\rm K}.
\end{array}\right.
\end{eqnarray}
If we assume that the eddy turn over frequency is equal to the Keplerian
frequency, $\omega_{\rm t}=\Omega_{\rm K}$, the fluctuation part
(\ref{eq19.2}) becomes $Z'(t)\approx(v_t/\cs)H\exp[i(\Omega_{\rm
K}t-\pi/2)]$, 
which means that 
the turbulent gas motion forces the
dust particle to continue to oscillate vertically with an amplitude
$(v_t/\cs)H$ and a frequency $\Omega_{\rm K}$, independent of the
dust particle size (e.g., Landau et al. 1967). 
If we think a more realistic case, the turbulent velocity of the
gas, $u_z'$, will be modeled by a superposition of eddies with various
sizes, velocities, and frequencies, which is often decomposed into
Fourier components (e.g., Landau \& Lifshitz 1959). In this case the
term for fluctuating motion (\ref{eq19.2}) 
is also given by a superposition of oscillations with various frequencies.
Now, the mean motion part
(\ref{eq19.1}) shows the settling of the dust particle toward the
disk midplane. From equation (\ref{eq19.0}) we can derive the particle
velocity, which also consists of a mean motion, $\overline{V}(t)$, and a
fluctuation, $V'(t)$, as
\begin{equation}
V(t)=\overline{V}(t)+V'(t),
\end{equation}
where $\overline{V}(t)=d\overline{Z}(t)/dt$ and $V'(t)=dZ'(t)/dt$.
For a small particle which satisfies $D_g>2\Omega_{\rm K}$, we obtain
$\overline{Z}(t)\simeq Z_0\exp[-(\Omega_{\rm K}^2/D_g)t]$ ($Z_0$ is the
initial value of $\overline{Z}(t)$) and $\overline{V}(t)\simeq
-(\Omega_{\rm K}^2/D_g)\overline{Z}(t)$.

We note that in a quiescent disk in which $u_z=0$, a general solution of
equation (\ref{eq17}) is simply given by the mean motion part,
$\overline{Z}(t)$, in equation (\ref{eq19.1}).
Therefore, the dust particles always settle toward the disk midplane; 
the motion of a large dust particle which satisfies $D_g<2\Omega_{\rm K}$
is oscillation around $Z=0$, damped (that is, the particle settles
toward the disk midplane) with a timescale of $2/D_g$, while
a smaller particle ($D_g>2\Omega_{\rm K}$) settles with a timescale
of $D_g/\Omega_{\rm K}^2$ without oscillation (e.g., NSH86).  


\subsection{Radial Motion}

Next, we will discuss the radial motion of the dust particles in a
turbulent disk. When the dust particles are small enough, their motion
is strongly coupled with the turbulent gas motion. If the particles grow
and reach
a critical radius, they are released from the turbulent eddy trapping and
migrate toward the central star (see e.g., Klahr \& Henning 1997
for more detailed description of the dust particle motion in a turbulent
eddy). The critical radius, $a_{\rm crit}$, is roughly estimated by comparing 
the friction time between the gas and dust particles, $\tau_f=1/D_g$,
with the turnover time of the largest turbulent eddy, 
$\tau_{\rm eddy}=1/\Omega_{\rm K}$, and given by
\begin{equation}
a_{\rm crit}=\left\{\begin{array}{ll}
\cs\rho_{\rm gas}/\rho_s\Omega_{\rm K} & {\rm for}\ \ a\la l_g, \\
(3\cs\rho_{\rm gas}l_g/2\rho_s\Omega_{\rm K})^{1/2} & {\rm for}\ \ a\ga l_g.
\end{array}\right.
\end{equation}
The critical radii at $R=$1, 5.2, and 30AU are $a_{\rm crit}=32$, 80,
and 6 cm, respectively. When the dust radius reaches $a_{\rm crit}$ and
the friction time, $\tau_f$, becomes as long as the eddy turnover time,
$\tau_{\rm eddy}$, the radial velocity of the particle becomes the
maximum, $V_R\simeq\eta v_{\rm K}=5\times 10^3$cm s$^{-1}$ (see
eq. [\ref{eq10}]), and the particle migrates inward very rapidly with
the timescales of $R/V_R\simeq 1\times 10^2, 5\times 10^2$, and $3\times
10^3$ yrs at $R=1$, 5.2, and 30AU, respectively 
(e.g., Adachi et al. 1976; Weidenschilling 1977).

\subsection{Dust Size Growth in a Turbulent Disk} \label{Sect4.3}

Next, taking into account the properties of vertical and radial motion
of dust particles mentioned in the previous subsections, we will
numerically simulate the dust evolution in a turbulent disk by solving
the coagulation equation (\ref{eq12}). 
In this simulation we artificially remove the dust particles whose radii
reach $a_{\rm crit}$ as they migrate inward very rapidly. Numerical
simulation including radial mass transport of the dust particles 
should be done in future (cf. Weidenschilling 2004).

As we mentioned in \S \ref{Sect2.3} the relative velocity between the
dust particles induced by microscopic motion of the turbulent gas,
$\delta V_t$, is taken into account in this numerical calculation. 
As the relative velocity we adopt the approximate treatment by
Weidenschilling (1984), which reproduces the result of V\"{o}lk 
et al. (1980)'s analysis of the nonlinear response of a dust particles
to the turbulent gas motion with a Kolmogorov spectrum.
In addition, we use Mizuno et al. (1988)'s formula when
the friction time is shorter than the turnover time of the smallest
turbulent eddy (see also Markiewicz et al. 1991). The adopted relative
velocity is 
\begin{equation}
\delta V_t=\left\{\begin{array}{ll}
\dfrac{3\tau_{f_j}}{(\tau_{f_i}+\tau_{f_j})}\biggl(\dfrac{\tau_{f_j}}{\tau_{k_0}}\biggr)^{1/2}v_t & {\rm for}\ \ \tau_{f_i} \leq \tau_{f_j} <\tau_{k_0}, \\
\biggl(\dfrac{|\tau_{f_i}-\tau_{f_j}|}{\tau_{f_i}+\tau_{f_j}}\biggr)^{1/2}\biggl|\dfrac{\tau_{f_i}}{\tau_{k_0}}\ln\dfrac{\tau_{k_0}+\tau_{f_i}}{\tau_{k_s}+\tau_{f_i}}-\dfrac{\tau_{f_j}}{\tau_{k_0}}\ln\dfrac{\tau_{k_0}+\tau_{f_j}}{\tau_{k_s}+\tau_{f_j}}\biggr|^{1/2}v_t & {\rm for}\ \ \tau_{f_i}, \tau_{f_j} \leq\tau_{k_s}, \label{eq22}
\end{array}\right.
\end{equation}
where $\tau_{f_i}$ is the friction time between the gas and dust
particles with a radius $a_i$, that is, $\tau_f=1/D_g$ for $a=a_i$.
The times of $\tau_{k_0}(=\tau_{\rm eddy}=1/\Omega_{\rm
K})$ and $\tau_{k_s}=Re^{-1/2}\tau_{k_0}$ are the turnover times of the
largest ($l_t$) and smallest ($Re^{-3/4}l_t$) turbulent eddies,
respectively. The Reynolds number, $Re$, is estimated as 
\begin{equation}
Re=v_tl_t/\nu=\alpha\cs H/\nu=2\times 10^{11}(R/{\rm AU})^{-3/2}\alpha,
\end{equation}
where we adopt the molecular viscosity of
$\nu=\cs l_g/3=(\cs m_{\mu})/(3\sigma_{\rm mol}\rho_{\rm gas}$) (e.g.,
Jeans 1916).
In this work we calculate the
dust evolution in a weakly and strongly turbulent disk, in which we
adopt $\alpha=10^{-4}$ ($v_t=10^{-2}\cs$ and $l_t=10^{-2}H$)
and $\alpha=10^{-2}$ ($v_t=10^{-1}\cs$ and $l_t=10^{-1}H$),
respectively. 

As the mass transport of dust particles in the vertical direction, we
take into account the transport due to turbulent
mixing as we mentioned in \S \ref{Sect2.3}. 
If we separate the particle mass density and the velocity into mean and
fluctuating parts, the mass flux
in the second term of the left hand side of equation (\ref{eq12}) is
described as $V_Z(i)\cdot\varphi(i)=\overline{V_Z(i)}\cdot
\overline{\varphi(i)}+\overline{V_Z'(i)\cdot\varphi'(i)}$,
where the overline means a time average and the prime is a fluctuation
due to the turbulent motion. As the mean vertical velocity we adopt
$\overline{V_Z(i)}=-(\Omega_{\rm K}^2/D_g)Z$ (see \S \ref{S4.1}). 
The second term of the right hand side of the equation,
$\overline{V_Z'(i)\cdot\varphi'(i)}$, is the correlation of the
fluctuations and treated as turbulent mixing, following the gradient
diffusion hypothesis; $\overline{V_Z'(i)\cdot\varphi'(i)}
=-D_0[\partial\overline{\varphi(i)}/\partial z]$, which works so as to
diffusively uniform the mass density gradient of $\varphi(i)$ (we omit
the overline hereafter) in the vertical direction 
as the dust particles move around from eddy to eddy. 
For the diffusivity, we adopt 
$D_0=v_tl_t/(1+\tau_f/\tau_{\rm eddy})=\alpha\cs H/(1+\Omega_{\rm
K}/D_g)$ (e.g., Cuzzi et al. 1993; Weidenschilling 1997).
The equation (\ref{eq12}) is, therefore, solved by adopting 
\begin{equation}
V_Z(i)\cdot\varphi(i)=-\dfrac{\Omega_{\rm K}^2}{D_g}Z\varphi(i)-D_0\dfrac{\partial\varphi(i)}{\partial z} 
\end{equation}
for simulating the dust evolution in a turbulent disk.

\begin{figure}
\plotone{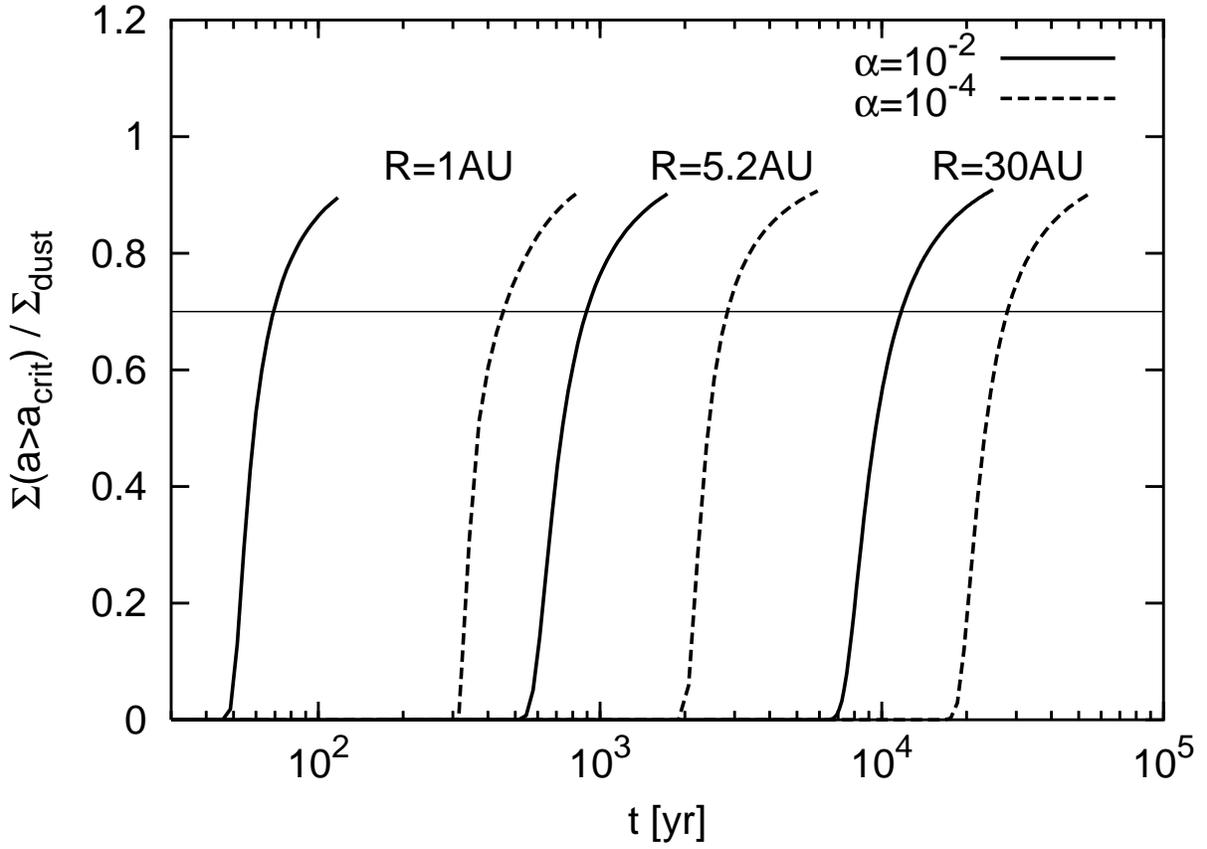}
\caption{The time evolution of the surface density of the dust particles
 which are larger than $a_{\rm crit}$ in strongly ($\alpha=10^{-2}$;
 solid lines) and weakly ($\alpha=10^{-4}$; dashed lines) turbulent
 disks. Most mass of dust particles is included in large particles,
 which can migrate toward the central star, at
 a very short timescale ($\sim 70$--$3\times 10^4$yr at 1--30 AU).
\label{f6}}
\end{figure}

In Figure \ref{f6} we plot the resulting time evolution of the surface
density of 
the dust particles whose radii reach $a_{\rm crit}$, $\Sigma(a>a_{\rm
crit})=\int_{-H}^{H}dZ\sum_{i=i_{\rm crit}}^n\varphi(i)$
($i_{\rm crit}$ corresponds to $a_{\rm crit}$), divided by the total
dust surface density, $\Sigma_{\rm dust}$ (eq.[\ref{eq2}]). 
As mentioned before, we have removed those large particles in the
numerical simulation, taking into account the rapid radially inward
migration. 
The figure shows that more than 70\% of the total dust mass moves
toward the central star very rapidly within about 70, $9\times 10^2$,
and $1\times 10^4$ yrs in a strongly turbulent disk ($\alpha=10^{-2}$;
solid lines), while about $5\times 10^2$, $3\times 10^3$, and $3\times
10^4$ yrs in a weakly 
turbulent disk ($\alpha=10^{-4}$; dashed lines), at the orbits of
$R=$1, 5.2, and 30 AU, respectively. The timescale of the dust size
growth is shorter at the inner disk where the particle density is higher. 

\begin{figure}
\begin{center}
\includegraphics[scale=1.0]{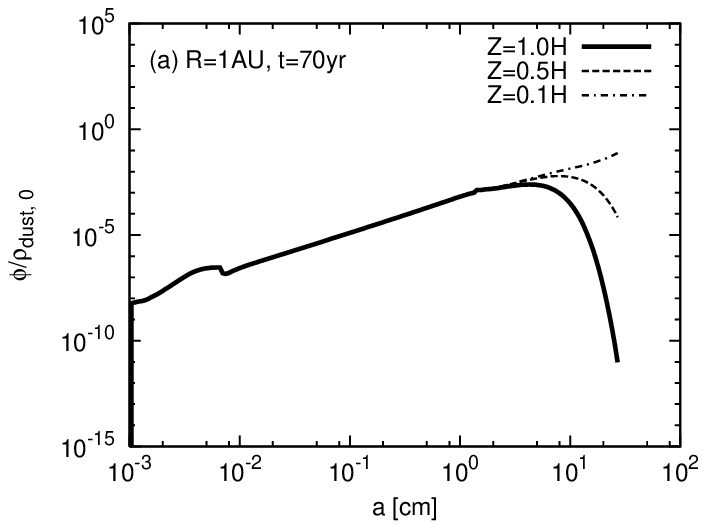}

\includegraphics[scale=1.0]{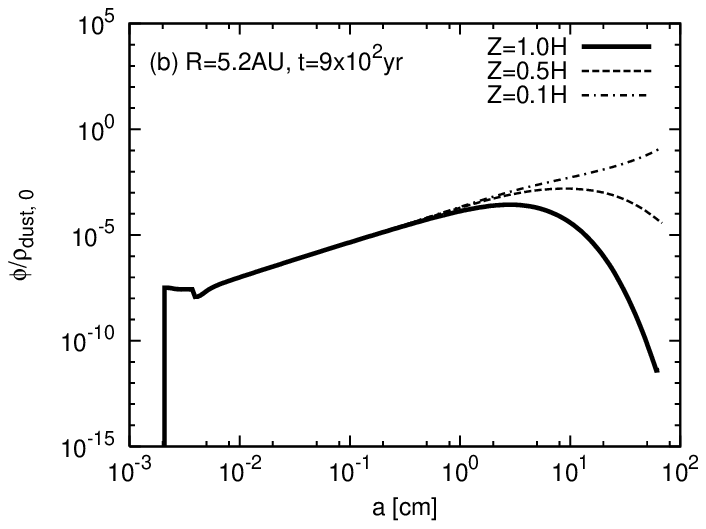}

\includegraphics[scale=1.0]{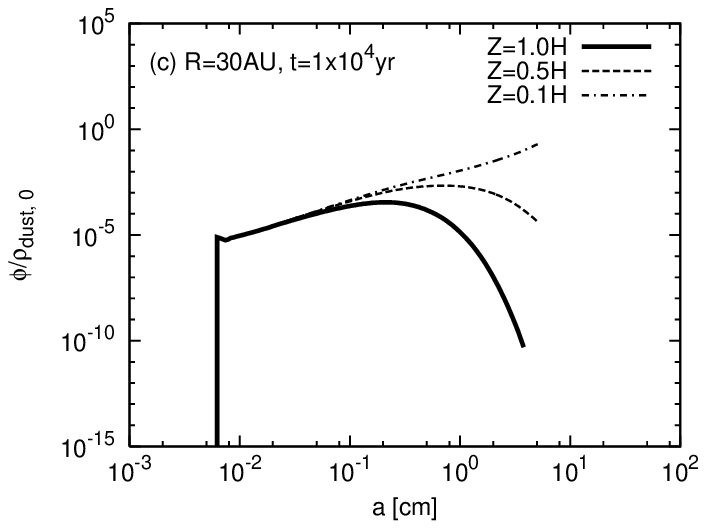}
\caption{The size distributions of mass density of dust particles,
 $\varphi(i)$, normalized by $\rho_{{\rm dust}, 0}$ at each orbit, $R$,
 and time, $t$ in a strongly turbulent disk ($\alpha=10^{-2}$). 
 Note that the normalization factor is different from Figure \ref{f1}
 (see text). 
 The thick solid, dashed, and dot-dashed lines represent the size
 distributions at $Z=H$, $0.5H$, and $0.1H$, respectively.
 The size distributions of smaller dust particles are almost
 identical at each height mainly because the turbulence induced motion
 dominates the relative velocity in almost all regions. Larger
 particles near the disk surface deplete since the turbulent diffusion
 against the settling toward the disk midplane is not strong due to
 weak coupling with the gas motion. \label{f5}}
\end{center}
\end{figure}

\begin{figure}
\begin{center}
\includegraphics[scale=1.0]{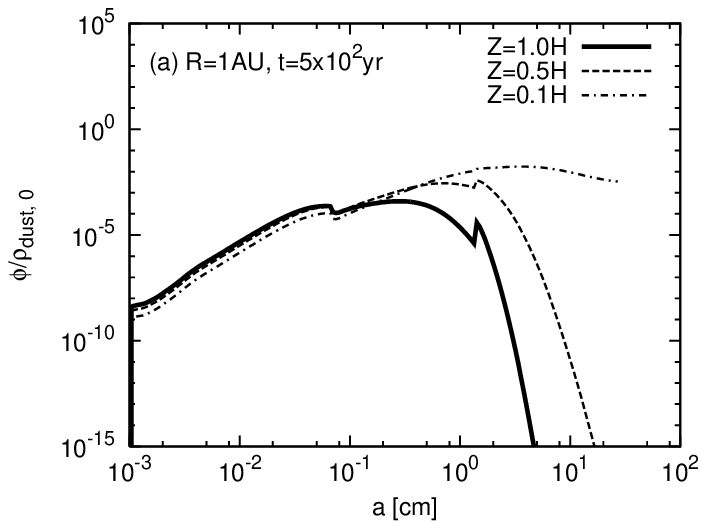}

\includegraphics[scale=1.0]{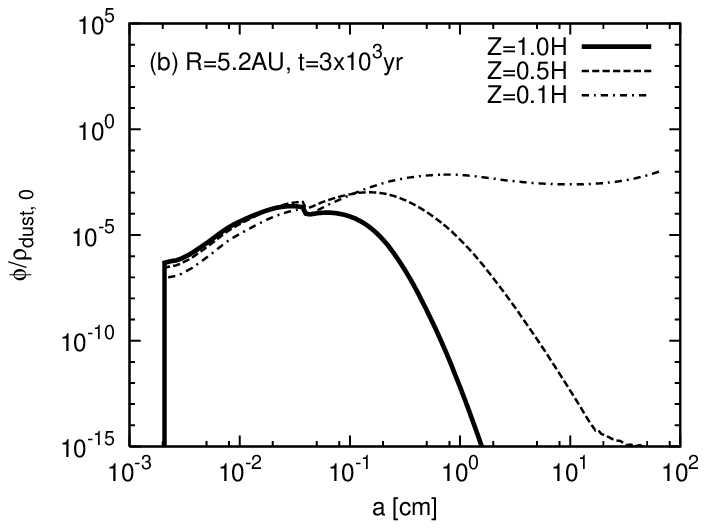}

\includegraphics[scale=1.0]{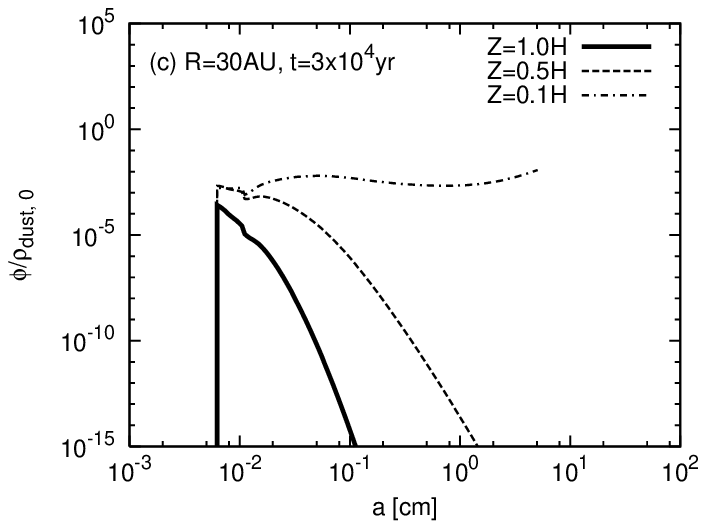}
\caption{The same as Figure \ref{f5}, but in a weakly
 turbulent disk ($\alpha=10^{-4}$). The size distributions of
 smaller dust particles are not very different at each height because
 of the turbulent mixing in the vertical direction. The depletion
 of larger particles at the upper disk layer is more remarkable than that
 in the strongly turbulent disk.
\label{f5.5}}
\end{center}
\end{figure}

Figure \ref{f5} shows the resulting size distributions of mass density
of dust particles, $\varphi(i)$, normalized by $\rho_{{\rm dust}, 0}$ at
(a) $R=1$AU, $t=70$ yr, (b) $R=5.2$AU, $t=9\times 10^2$ yr, and (c)
$R=30$AU, $t=1\times 10^4$ yr, in a strongly turbulent disk
($\alpha=10^{-2}$). 
Figure \ref{f5.5} is the same as Figure \ref{f5}, but in a
weakly turbulent disk ($\alpha=10^{-4}$), at (a) $R=1$AU, $t=5\times
10^2$ yr, (b) $R=5.2$AU, $t=3\times 10^3$ yr, and (c) $R=30$AU,
$t=3\times 10^4$ yr. We note that the normalization factor in
Fiugres \ref{f5} and \ref{f5.5} is different from that in Figure \ref{f1};
$\rho_{{\rm dust}, 0}$ used in Fiugres \ref{f5} and \ref{f5.5} is the
initial dust density, $\rho_{{\rm dust}, 0}=4.2\times 10^{-3}\rho_{\rm
gas}$ and $1.8\times 10^{-2}\rho_{\rm gas}$ for $R<2.7$AU and $R>2.7$AU,
respectively, while $\rho_{\rm dust}$ used in Figure \ref{f1} is the
dust density at a specific time and spatial position defined in equation
(\ref{eq14}). The thick solid, dashed, and dot-dashed lines
in each figure represent the size distributions at $Z=H$, $0.5H$, and
$0.1H$, respectively. 
The time used in the figures is when 70\% of the dust particles grow
large enough to migrate toward the central star very rapidly at each
orbit, $R$ (see Fig.~\ref{f6}). 
We can see from the figures that at each orbit, $R$, the size
distributions of smaller dust particles are almost identical at
each height, while those of larger particles are very different. 
In the strongly turbulent disk ($\alpha=10^{-2}$; Fig.~\ref{f5})
the smaller particles have similar size distributions at each height
mainly because the turbulence 
induced motion dominates the relative velocity and the dust size growth
(e.g., Weidenschilling 1984) in almost all disk heights, $Z$; i.e.,
$\delta V\simeq \delta V_t$ independent of $Z$. 
%
Meanwhile, in the weakly turbulent disk ($\alpha=10^{-2}$;
Fig.~\ref{f5.5}) the differential vertical
velocity, $\delta V_Z$, dominates the relative velocity at $Z\approx H$
where the gravitational force in the vertical direction is strong, while
the turbulence induced relative velocity, $\delta V_t$, is dominant near
the disk midplane.
Therefore, the dust particles grow more
rapidly at the disk surface, $Z\approx H$, 
and small particles
are replenished from lower disk layers via the turbulent mixing,
which works so as to uniform $\varphi(i)$. Consequently, the size
distributions of smaller particles are not very different at each
height also in the weakly turbulent disk. While the diffusive
motion of turbulent mixing is strong enough to prevent the settling for
the smaller particles, the larger particles cannot be sustained because
of weak coupling with the gas and settle toward the disk midplane. So, the
larger particles near the disk surface deplete as we can see from the 
figures. The depletion is more remarkable in the weakly turbulent disk
(e.g., Dubrulle et al. 1995; Cuzzi et al. 1996; Cuzzi \& Weidenschilling
2005).
The gaps appear around $a=$(7, 4, and 1.1)$\alpha^{-1/2}\micron$ at
$R=1$, 5.2, and 30AU, respectively, in Figures \ref{f5} and \ref{f5.5},
owing to the discontinuities of the approximate treatment of the
turbulence induced relative motion at $\tau_f=\tau_{k_s}$ (see
eq. [\ref{eq22}]).

Our result that the most dust particles migrate toward the central star
at a very short timescale suggests that global turbulent motion should
cease for the planetesimal formation in protoplanetary disks. 
Unless the strength of turbulence is weak, the density of dust particles
around the disk midplane will be low enough because of the turbulent
stirring so that they cannot collisionally grow into planetesimals, as
has been noted also by some previous works (e.g., Stepinski \& Valageas
1996; Cuzzi \& Zahnle 2004; Weidenschilling 2004; Cuzzi \&
Weidenschilling 2005). In a quiescent disk the
dust particles will settle toward the disk midplane and form a dust
layer in a sufficiently short timescale as we have seen in the previous
section. Afterwards the shear between the dust layer and the gas will
cause a local turbulent motion near the midplane. Detailed analysis of
the evolution of dust particles in such a turbulent dust layer, although
it is beyond a scope of this work, is in progress for further
understanding of the planetesimal formation process (e.g., Goldreich \&
Ward 1973; Cuzzi et al. 1993; Weidenschilling 1995; Sekiya 1998; Ishitsu
\& Sekiya 2003; 
Youdin \& Shu 2002; Weidenschilling 2003; Youdin \& Chiang 2004).

\section{Summary}

We have investigated the dust size growth and settling toward the disk
midplane in a quiescent or turbulent protoplanetary disk by numerically
solving coagulation equation for settling dust particles.

Our result shows that the dust particles settle toward the disk
midplane to form a gravitationally unstable layer at a short timescale
($2\times 10^3$--$4\times 10^4$yr at $R=$1--30 AU) if we assume
an ideally quiescent disk. 
The radii of the largest dust particles just before the formation of the
unstable layer are 20--1 cm at 1--30 AU.
The resulting settling time and evolution of the largest dust radius in
our numerical simulation are in good agreement with those obtained by the
analytic calculation in NSH86, although they did not take into
account the dust size distribution explicitly. This is
because most dust mass is included in the dust particles with the
largest sizes, and these particles control the dust settling time.

Also, we have discussed the dust evolution in an opposite extreme
case of a globally turbulent disk to find that the dust particles 
are forced to fluctuate by turbulent motion of the gas,
and grow to be large enough (32--6 cm
at 1--30AU) to move inward very rapidly within a short timescale 
($70$--$3\times 10^4$yr at 1--30 AU). Thus, our result suggests 
that the disk should be quiescent or the global turbulent motion should
cease before most mass of dust particles accrete onto the central star,
in order to form planetesimals in protoplanetary disks. 
Self-consistent treatment of the evolution of the globally turbulent
regions and the dust evolution processes is needed in future work.
In addition, the dust evolution in a locally turbulent motion induced by
the shear between the dust layer and the gas near the disk midplane
should be investigated for further understanding of the planetesimal
formation process.

\acknowledgments

We would like to thank the referee, Dr. S.J. Weidenschilling for
his comments which were greatly helpful in improving our paper. 
Also, we are grateful to Makoto Kohno for arranging the numerical code
for coagulation of settling dust particles.
This work is supported by ``The 21st Century COE Program of Origin and
Evolution of Planetary Systems" and the Grants-in-Aid for
Scientific Research 17540217, 17039008, and 16036205 in MEXT.






\appendix

\section{Testing Numerical Solution of Coagulation Equation}

It is known that a numerical calculation of coagulation equation with
inappropriate conditions, for example, coarse mass bins, causes some
serious errors like an artificial acceleration of coagulation that
could lead to an artificial runaway (e.g., Ohtsuki et al. 1990; Wetherill
1990). Here we test our numerical solution of coagulation equation by
comparing it with an analytic solution, using different numerical
conditions. A linear kernel of $m_im_j\beta(i,j)=Q(m_i+m_j)$ ($Q$ is a
constant) is used as a coalescence rate. In this case the coagulation
equation with an initial condition of
$\varphi(i)/(m_i/m_1)^2|_{t=0}=\rho_{\rm dust}e^{-m_i/m_1}$ 
has an analytic solution, 
\begin{eqnarray}
\dfrac{\varphi(i)}{(m_i/m_1)^2} &=& \dfrac{\rho_{\rm dust}\exp[-\eta+(m_i/m_1)(2-e^{-\eta})]}{(m_i/m_1)(1-e^{-\eta})^{1/2}}I_1[2(m_i/m_1)(1-e^{-\eta})^{1/2}] \\
 &\approx& \dfrac{\rho_{\rm dust}\exp\{-\eta-(m_i/m_1)[1-(1-e^{-\eta})^{1/2}]^2\}}{2\pi^{1/2}(m_i/m_1)^{3/2}(1-e^{-\eta})^{3/4}} \label{eqA2}
\end{eqnarray}
where $I_1(x)$ is the modified Bessel function and $\eta=Q\rho_{\rm
dust}t$  (Safronov 1963, 1969). 

\begin{figure}
\begin{center}
\includegraphics[scale=1.0]{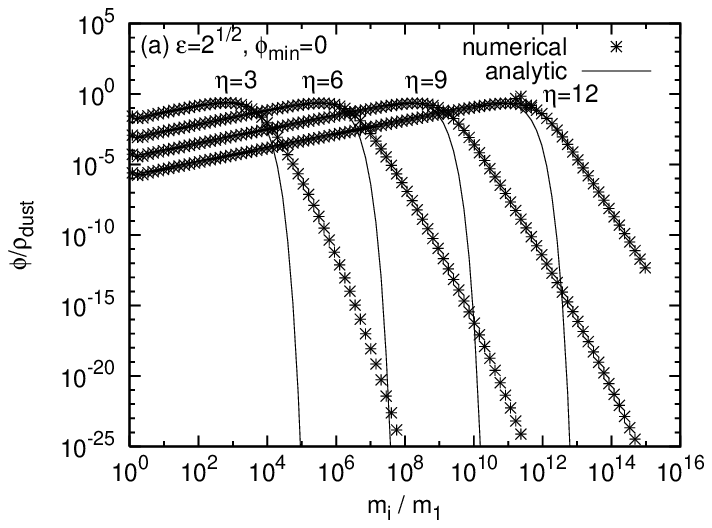}

\includegraphics[scale=1.0]{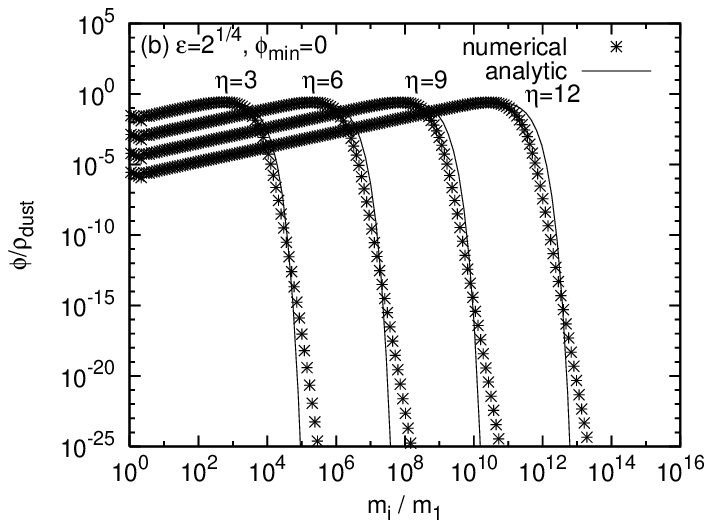}

\includegraphics[scale=1.0]{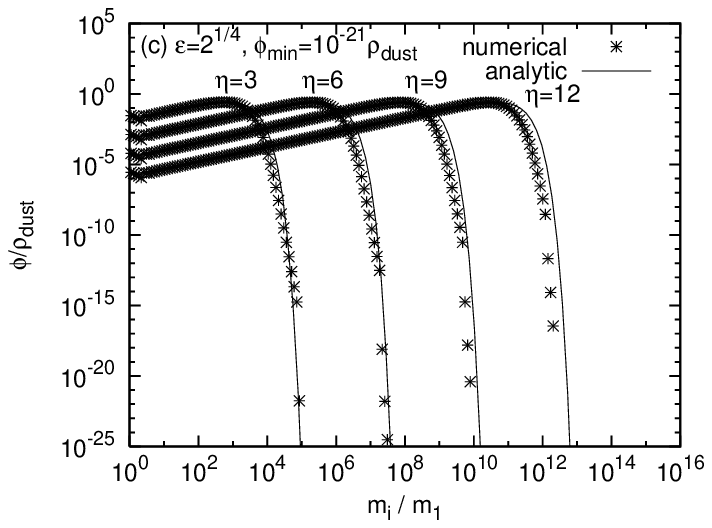}
\caption{Analytic (solid lines) and numerical (crosses) solutions for
 the mass distributions at $\eta=$ 3, 6, 9, and 12. Different numerical
 conditions are used in 
each figure; (a) $\varepsilon=\sqrt{2}$, $\varphi_{\rm min}=0$,
(b) $\varepsilon=\sqrt[4]{2}$, $\varphi_{\rm min}=0$, and (c)
$\varepsilon=\sqrt[4]{2}$, $\varphi_{\rm min}=10^{-21}\rho_{\rm
 dust}$. The conditions $\varepsilon=\sqrt[4]{2}$ and 
$\varphi_{\rm min}=10^{-21}\rho_{\rm dust}$ are used in this
 paper. \label{f7}} 
\end{center}
\end{figure}

In Figure \ref{f7} we compare the analytic solutions (\ref{eqA2})
(solid lines) and the numerical solutions (crosses) which have an
initial condition for the dust mass distributing at
$m_1$. The mass distributions divided by $\rho_{\rm dust}$ at $\eta=$ 3,
6, 9, and 12 are plotted. Different numerical conditions are used in
each figure; (a) $\varepsilon=\sqrt{2}$, $\varphi_{\rm min}=0$,
(b) $\varepsilon=\sqrt[4]{2}$, $\varphi_{\rm min}=0$, and (c)
$\varepsilon=\sqrt[4]{2}$, $\varphi_{\rm min}=10^{-21}\rho_{\rm dust}$,
where $\varepsilon$ is the intervals of the dust mass
bins, $m_{i+1/2}=\varepsilon m_{i-1/2}$ (see \S\ref{Sect2.3}), and we
prohibit the collisional coagulation ($\beta(i,j)=0$) if
$\varphi(i)<\varphi_{\rm min}$ or $\varphi(j)<\varphi_{\rm min}$ (e.g.,
Ohtsuki et al. 1990). 
In this paper we choose the conditions $\varepsilon=\sqrt[4]{2}$ and
$\varphi_{\rm min}=10^{-21}\rho_{\rm dust}$, with which the numerical
solution is unlikely to cause the artificial runaway.




\clearpage




\clearpage


\clearpage

\clearpage

\clearpage

\clearpage

\clearpage

\clearpage

\clearpage

\begin{deluxetable}{cccc}
\tablecaption{Characteristic Disk Heights \label{T1}}
\tablewidth{0pt}
\tablehead{
\colhead{} & \colhead{$R=$1AU} & \colhead{$R=$5.2AU} & \colhead{$R=$30AU}}
\startdata
$Z_1/H$ & $7.7\times 10^{-2}$ & $1.2 \times 10^{-1}$ & $1.8 \times 10^{-1}$ \\
$Z_2/H$ & $3.7\times 10^{-3}$ & $1.6 \times 10^{-2}$ & $1.6 \times 10^{-2}$ \\
$Z_3/H$ & $1.8\times 10^{-4}$ & $2.1 \times 10^{-3}$ & $1.4 \times 10^{-3}$ \\
$Z_4/H$ & $2.4\times 10^{-6}$ & $1.5 \times 10^{-5}$ & $2.4 \times 10^{-5}$ \\
$H$ [AU] & $4.7\times 10^{-2}$ & $3.7 \times 10^{-1}$ & $3.3$ \\
\enddata
\end{deluxetable}


\end{document}